\documentclass[preprint,12pt]{elsarticle}

\usepackage{amssymb}
\usepackage{amsmath}
\usepackage{graphicx}

\journal{Physics Letters B}

\begin{document}

\begin{frontmatter}



\title{Measurement of $^{73}$Ge($n,\gamma$) cross sections and implications for stellar nucleosynthesis}

\author[a30,a9]{C.~Lederer-Woods\corref{cor1}}\cortext[cor1]{Corresponding author: claudia.lederer-woods@ed.ac.uk}
\author[a30]{U.~Battino}
\author[a7]{P.~Ferreira}
\author[a2]{A.~Gawlik}
\author[a15]{C.~Guerrero}
\author[a10,a1]{F.~Gunsing}
\author[a21]{S.~Heinitz}
\author[a15]{J.~Lerendegui-Marco}
\author[a33]{A.~Mengoni}
\author[a9]{R.~Reifarth}
\author[a30]{A.~Tattersall}
% ...
\address[a30]{School of Physics and Astronomy, University of Edinburgh, United Kingdom }
\address[a9]{Goethe University Frankfurt, Germany }
\address[a7]{Instituto Superior T\'{e}cnico, Lisbon, Portugal }
\address[a2]{University of Lodz, Poland }
\address[a15]{Universidad de Sevilla, Spain }
\address[a10]{CEA Irfu, Universit\'{e} Paris-Saclay, F-91191 Gif-sur-Yvette, France }
\address[a1]{European Organization for Nuclear Research (CERN), Switzerland }
\address[a21]{Paul Scherrer Institut (PSI), Villingen, Switzerland }
\address[a33]{Agenzia nazionale per le nuove tecnologie, l'energia e lo sviluppo economico sostenibile (ENEA), Bologna, Italy }
% ...
\begin{abstract}
$^{73}$Ge($n,\gamma$) cross sections were measured at the neutron time-of-flight facility n\_TOF at CERN up to neutron energies of 300~keV, providing for the first time experimental 
data above 8~keV. Results indicate that the stellar cross section at $kT=30$~keV is
1.5 to 1.7 times higher than most theoretical predictions. The new cross sections result in a substantial decrease of $^{73}$Ge produced in stars, which would explain the low isotopic abundance of $^{73}$Ge in
the solar system. 
\end{abstract}

\begin{keyword}
nucleosynthesis \sep neutron capture \sep $s$ process \sep germanium \sep n\_TOF



\end{keyword}

\end{frontmatter}



\section{Introduction}
About half of the chemical element abundances heavier than iron in our solar system are produced by the slow neutron capture process ($s$ process) in stars.
The $s$ process takes place at moderate neutron densities around $10^{8}$ cm$^{-3}$, where neutron captures and subsequent radioactive $\beta$ decays build up the isotopes along the line of stability. 
The $s$ process in massive stars (so-called weak component) is mainly responsible for forming elements between Fe and Zr \cite{PET68,COUCH74,LAMB77,RAIT91a,RAIT91b}. 
In this scenario, neutrons are produced by $^{22}$Ne($\alpha,n$) reactions in two different burning stages, first during He core burning at temperatures of 0.3 GK (GK=$10^{9}$ K),
and later during carbon shell burning at 1 GK temperature. Solar germanium is thought to be mainly produced by the weak $s$ process (Pignatari et al. estimate 80\% \cite{PIG10}),
with the remaining contributions coming from the $s$ process in Asymptotic Giant Branch stars (main component), and explosive nucleosynthesis. 
Neutron capture cross sections averaged over the stellar neutron energy distribution (Maxwellian Averaged Cross Sections)
are a key input to predict abundances produced in the $s$ process, and the isotopic abundance distribution of Ge is highly sensitive to neutron capture cross sections on germanium. 
The sensitivity study by \cite{Nishi17} found an especially large uncertainty for the $^{73}$Ge production in massive stars, with
$^{73}$Ge($n,\gamma$) being the key rate responsible for the uncertainty.\\ 
Present experimental data on $^{73}$Ge+n reactions are scarce. 
In the astrophysical energy range, capture and transmission data by Maletski et al. \cite{Mal68} provide radiative widths $\Gamma_\gamma$ for resonances up to 2 keV, and neutron widths $\Gamma_n$
up to 8~keV. However, this energy region contributes only to a small extent to the relevant stellar cross sections at $kT=26$ and $kT=90$ keV which is 
equivalent to 0.3 and 1 GK $s$-process temperatures in massive stars, respectively. 
In addition Harvey and Hockaday \cite{HH80} measured total cross sections on natural germanium for neutron energies up to 180~keV. These two datasets currently form the experimental basis
for evaluated cross section libraries such as ENDF/B-VIII \cite{ENDFVIII}. In this letter, we report for the first time $^{73}$Ge($n,\gamma$) cross sections up to 300 keV neutron energy. 
This measurement is part of a wider campaign to measure ($n,\gamma$) cross sections on all stable germanium isotopes at n\_TOF. \\
\section{Measurement}
The measurement was performed at the neutron time-of-flight facility n\_TOF at CERN. At n\_TOF, an intense neutron beam is produced by spallation reactions of a 20 GeV/c proton beam 
of the CERN Proton Synchrotron (PS), 
impinging on a massive lead target. The initially highly energetic neutrons are moderated with borated water, resulting in a neutron spectrum  which ranges from 25 meV to several GeV of energy. 
Further details about the n\_TOF facility can be found in Ref. \cite{guerrero2012}. The radiative capture measurement was performed at Experimental Area 1 (EAR1)
located at a distance of 185 m from the spallation target. The long distance from the spallation target combined with the 7 ns width of the PS proton beam results in a high neutron energy resolution ranging from
$\Delta E_n / E_n=3\times10^{-4}$ at 1 eV, to $\Delta E_n / E_n=3\times10^{-3}$ at 100 keV \cite{guerrero2012}. 
The prompt $\gamma$ rays emitted after the capture event were detected using a set of liquid scintillation detectors (C$_6$D$_6$). These detectors have a low sensitivity to neutrons and thus minimise 
background produced by neutrons scattered from the sample. The capture sample consisted of 2.69 g GeO$_{2}$ which was $96.1\%$ enriched in $^{73}$Ge \footnote{Enriched material in metal form was
not available from the supplier. The sum of all chemical impurities in this sample was quoted as $<200$ ppm by the supplier.}. The GeO$_{2}$ material was originally obtained in
powder form and was pressed into a self supporting cylindrical pellet of 2 cm diameter and a thickness of 2.9 mm. 
In addition to the GeO$_{2}$ pellet, we recorded neutron capture data with a Au sample of the same diameter for normalisation of the data, an empty sample holder for background measurements, and a metallic Ge sample 
of natural isotopic composition. The latter was used to unambiguously identify resonances due to other Ge isotopes and to confirm the stoichiometry of the pellet.

\section{Data Analysis and Results}
The neutron time-of-flight spectra were converted to neutron energy by determining the effective flight path using low energy resonances in Au, for which the resonance energy has been determined with high precision 
at the time-of-flight facility GELINA \cite{MBK11}.
The neutron capture yield at neutron energy $E_n$, defined here as the probability for a neutron to be captured in the sample, can then be determined as:
\begin{equation}
Y(E_n)=f_N(E_n)\frac{C(E_n)-B(E_n)}{\Phi(E_n)\epsilon_c},
\end{equation}
where $C(E_n)$ is the number of counts, $B(E_n)$ are counts due to background, and $\Phi(E_n)$ is the neutron flux spectrum. $\epsilon_c$ is the efficiency to detect a capture event and $f_N$ is a normalisation
factor (see below).  \\
The detection efficiency was taken into account using the Pulse Height Weighting Technique (PHWT) \cite{pwht1,pwht2}, which can be applied to low efficiency systems, where typically only one $\gamma$-ray per 
capture cascade is detected.  If the efficiency to detect a $\gamma$ ray is proportional to the $\gamma$-ray energy ($\epsilon_{\gamma}\propto E_{\gamma}$),  the efficiency to detect a capture event is 
proportional to the excitation energy of the compound nucleus, i.e. $\epsilon_{c} \propto \sum \epsilon_{\gamma}=S_n+E_\text{cm}$. 
The $\epsilon_{\gamma}\propto E_{\gamma}$ proportionality can be achieved by applying pulse height dependent
weights to each recorded event. The weighting factors were determined by simulating the detector response in GEANT4 \cite{geant4}, taking into account the geometry of the setup and the capture samples used. 
The data further need to be corrected for transitions without $\gamma$-ray emission (electron conversion) and the missing contribution of $\gamma$-rays with energies below the detection threshold,
which was set in the analysis to 350 keV. 
These contributions were estimated and corrected for by simulating capture cascades with the code DICEBOX \cite{Bec98}, which generates individual levels and their decay properties based on
existing experimental information below an excitation energy of 2.6 MeV, and is based on level densities and photon strength functions above.
The systematic uncertainty of the PHWT is 2\% \cite{pwht2}, taking into account the additional threshold corrections we assign 3\% systematic uncertainty in total.  \\
The background $B(E_n)$ consists of three components: (i) Background unrelated to the neutron beam, for example due to natural radioactivity, is determined in runs without neutron beam; (ii) Beam related background
is determined by a measurement without the Ge sample in the beam (empty sample holder);  (iii) Background related to the sample, for example due to neutrons scattered off the sample which are
captured somewhere else in the experimental area after a time delay. Component (iii) can be estimated using neutron filters. These filters are made of material which show strong neutron absorption resonances at certain energies. 
The thickness is chosen such, that neutron transmission at these energies is negligible. Any counts in the dips of these resonances therefore must be produced by background reactions.
Components (i) and (ii) were measured and subtracted from the counting spectra. Component (iii) is most important
at higher neutron energies, where individual resonances start to overlap due to the experimental resolution and to the widening of the resonance widths, and consequently, the signal to background ratio 
in the resonance decreases. 
This background was estimated by subtracting an empty sample holder spectrum with an Al filter from the $^{73}$Ge with Al filter measurement. 
The remaining counts in the filter dips due to resonances in Al (35, 90, 120 and 140~keV)
were considered to be due to background. A smooth function was fitted to these filter dips and subtracted from the $^{73}$Ge data. Due to the low statistics in the filter dips the uncertainty in the background 
level is 20-30\%, which translates into an uncertainty in the capture yield of at most 1\%. \\
The neutron flux was measured in a dedicated campaign using reactions with well known cross sections and three different detection systems to minimise systematic uncertainties. 
The flux measurement was performed with a set of silicon detectors using $^{6}$Li($n,t$) reactions (SiMon detector), a Micromegas detector measuring $^{6}$Li($n,t$) and $^{235}$U($n,f$), 
and an ionisation chamber provided by Physikalisch Technische Bundesanstalt Braunschweig, measuring $^{235}$U($n,f$). The data were then combined to produce a reliable flux over the entire neutron energy range. 
The final evaluated neutron flux has a systematic uncertainty below 1\% for neutron energies $<3$~keV, and of 3.5\% between 3~keV 
and 1~MeV \cite{BS16}. More details on the neutron flux evaluation at n\_TOF can be found in Ref. \cite{Barb13}. The neutron fluence was monitored throughout the measurement by recording the number
of protons impinging on the spallation target (provided by PS detectors). This was cross checked using the SiMon detector which was operational throughout the run. In addition,
the stability of the C$_6$D$_6$ detectors was monitored by integrating the total number of counts in a strong $^{73}$Ge+n resonance at 103 eV for each run. No deviations outside statistical fluctuations were found. \\
The normalisation factor $f_N$ accounts for the fact that the neutron beam is larger than the capture sample, and corrects any inaccuracies in the solid angle coverage of the detectors assumed in simulations. 
This normalisation is determined using the saturated resonance technique on a $^{197}$Au$+n$ resonance at 4.9 eV neutron energy. For this resonance, the capture width is much larger than the neutron width 
which means that almost 100\% of neutrons interacting inside the sample are captured eventually. If the sample is chosen sufficiently thick, it can be ensured that all neutrons
traversing the sample react and  produce a $\gamma$-cascade, thus providing a direct measure of the neutron flux. The neutron beam size and hence the normalisation factor have a slight dependence on neutron energy. 
This energy dependence was determined in simulations \cite{guerrero2012} and corrections to the yield were at most $1.5\%$ in the investigated neutron energy range. 
The uncertainty assigned to the normalisation procedure is 1\%.\\
\begin{figure}[!htb]
\includegraphics[width=10.0 cm]{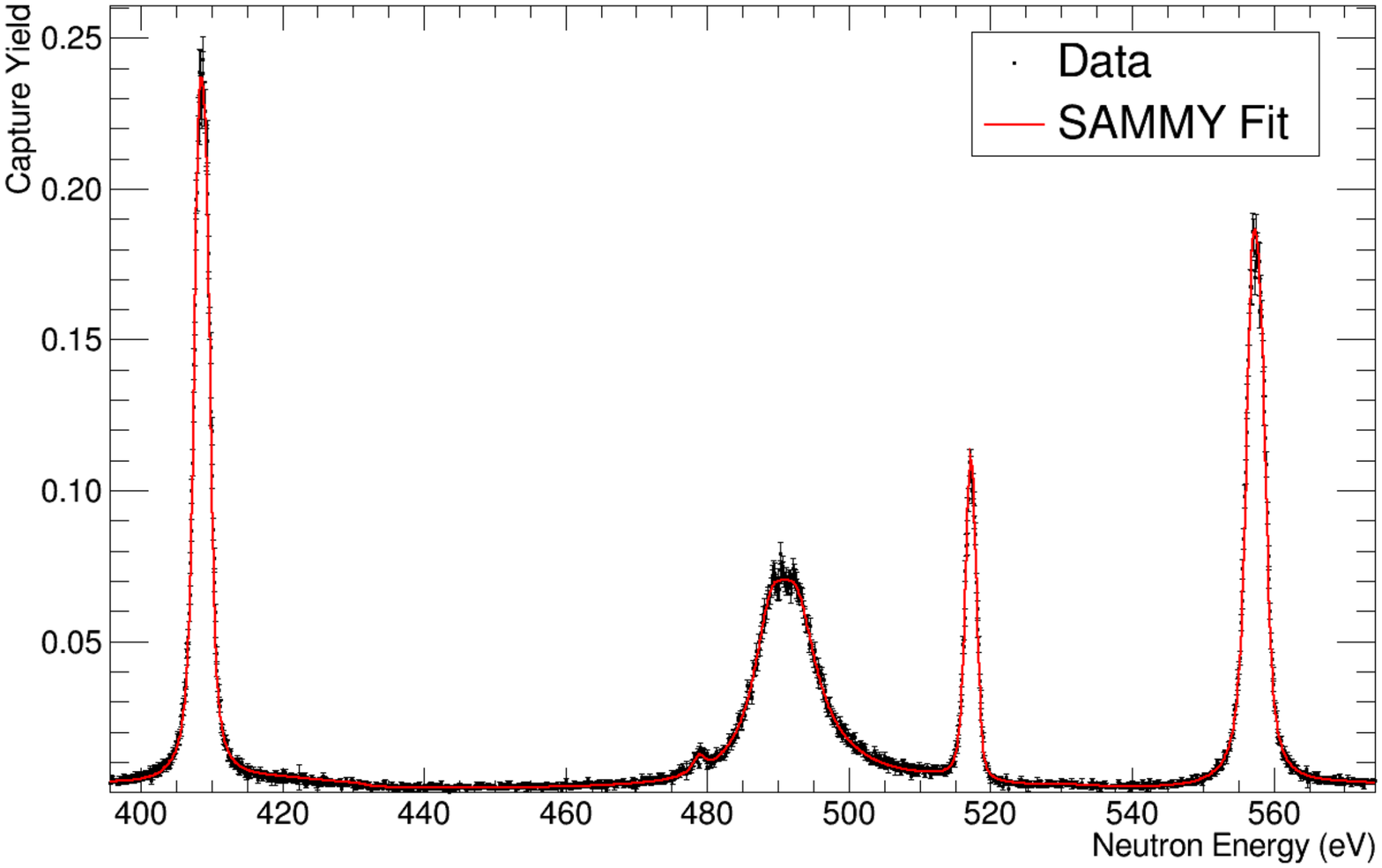}
\includegraphics[width=10.0 cm]{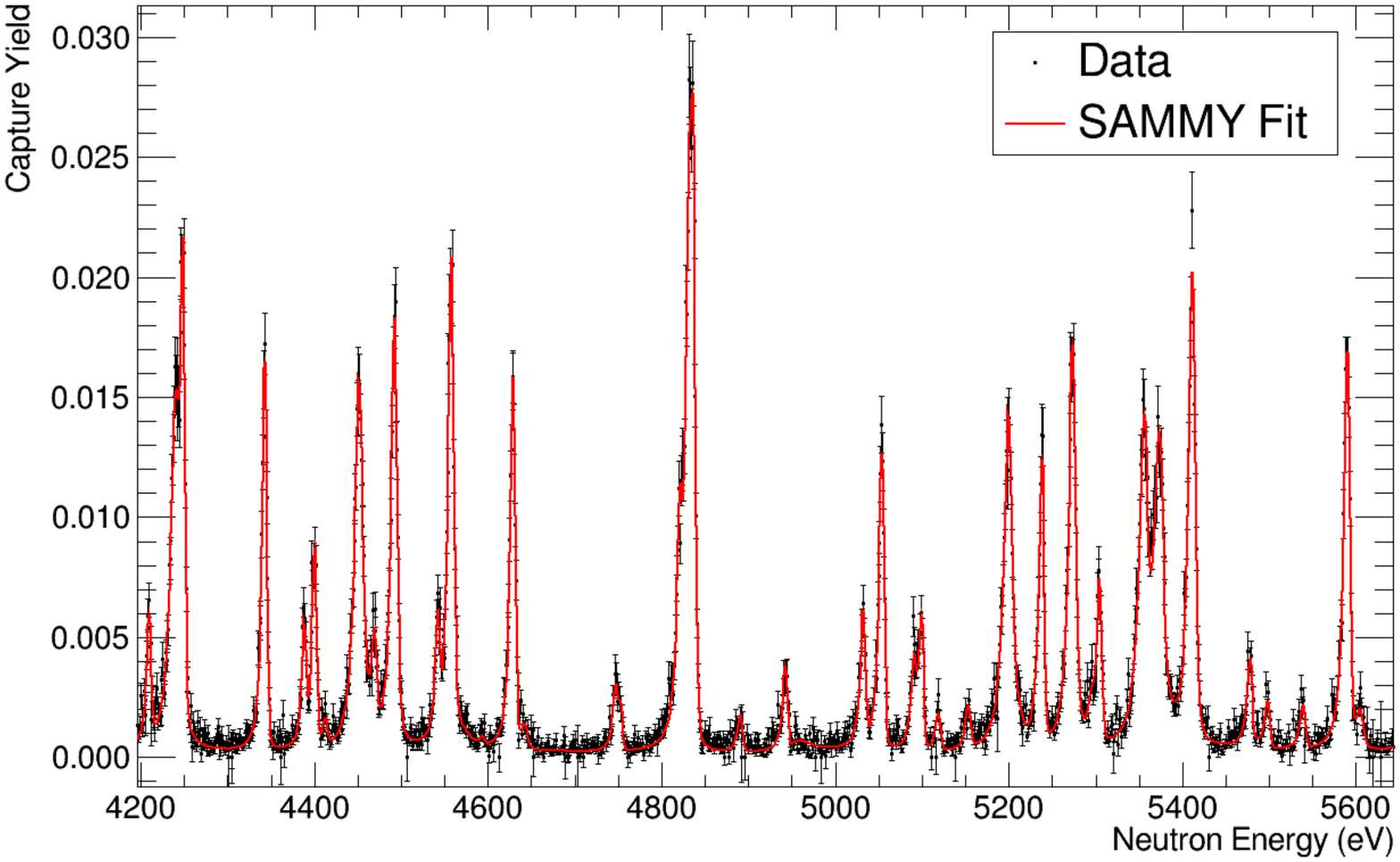}
\caption{Examples of SAMMY fits of the $^{73}$Ge($n,\gamma$) neutron capture yield (probability of a neutron capture)  obtained with the enriched $^{73}$GeO$_2$ sample for two different neutron energy regions. \label{resplots}}
\end{figure}
The resulting capture yield was fitted with the code SAMMY \cite{sammy}, a multilevel, multichannel R-matrix code using Bayes' method. SAMMY takes into account
experimental effects such as Doppler and resolution broadening, self shielding and multiple scattering of neutrons in the sample. SAMMY was used to extract resonance capture kernels $k$ up to 14 keV neutron energy:
\begin{equation}
k=g\frac{\Gamma_\gamma \Gamma_n}{\Gamma_\gamma+\Gamma_n},
\end{equation}
 where $g$ denotes the statistical weighting factor ($\frac{2J+1}{(2I+1)(2s+1)}$ with $J$ being the resonance spin, $I$ the target nucleus spin and $s$ the neutron spin), $\Gamma_\gamma$ is the radiative width, 
 and $\Gamma_n$ is the neutron width.  The list of resonance energies and capture kernels up to 14~keV can be found in \ref{kerneltables}. For a few low energy resonances, fitting the natural germanium sample
 resulted in a much better reproduction of the resonance shape, presumably due to its lower thickness which results in smaller corrections for multiple scattering and self shielding
in large resonances. The resonances fitted with the natural germanium samples are clearly marked in the list of resonances of Table \ref{kernels}. Examples of the SAMMY fits of the capture yield are shown in 
Figure \ref{resplots}.

\begin{figure}[!htb]
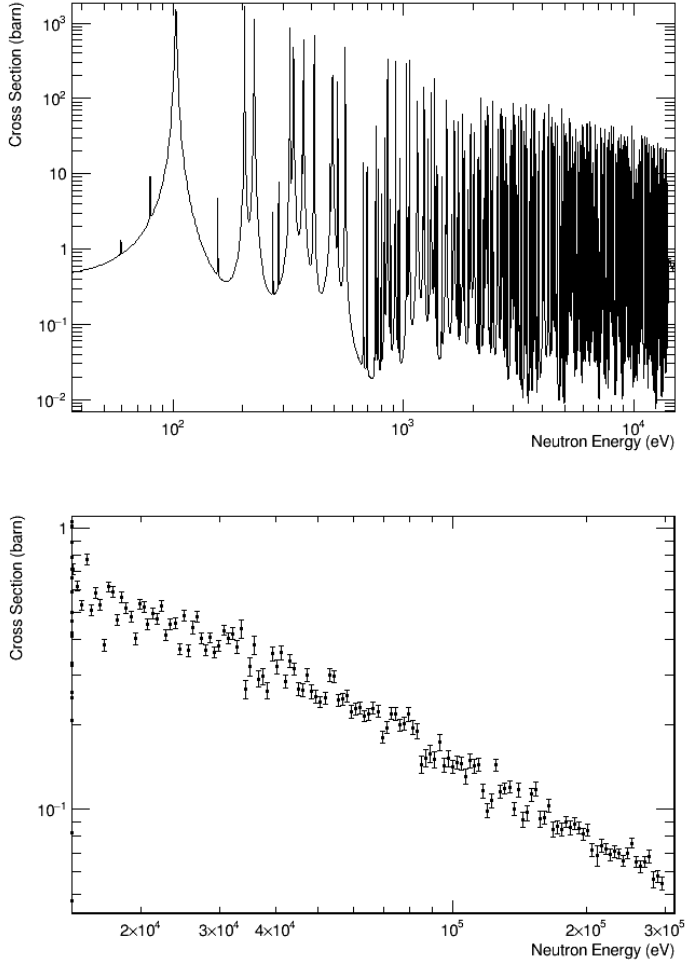

\includegraphics[width=10.0 cm]{URR_xs.pdf}
\includegraphics[width=10.0 cm]{URR_xs_corr.pdf}
\caption{(Top) $^{73}$Ge($n,\gamma$) cross sections reconstructed from SAMMY resonance fits up to 14~keV neutron energy. (Bottom) Averaged cross sections from 14~keV to 300~keV neutron energy
and statistical uncertainties. \label{xsplots}}
\end{figure}

\begin{figure}[!htb]
\includegraphics[width=10.0 cm]{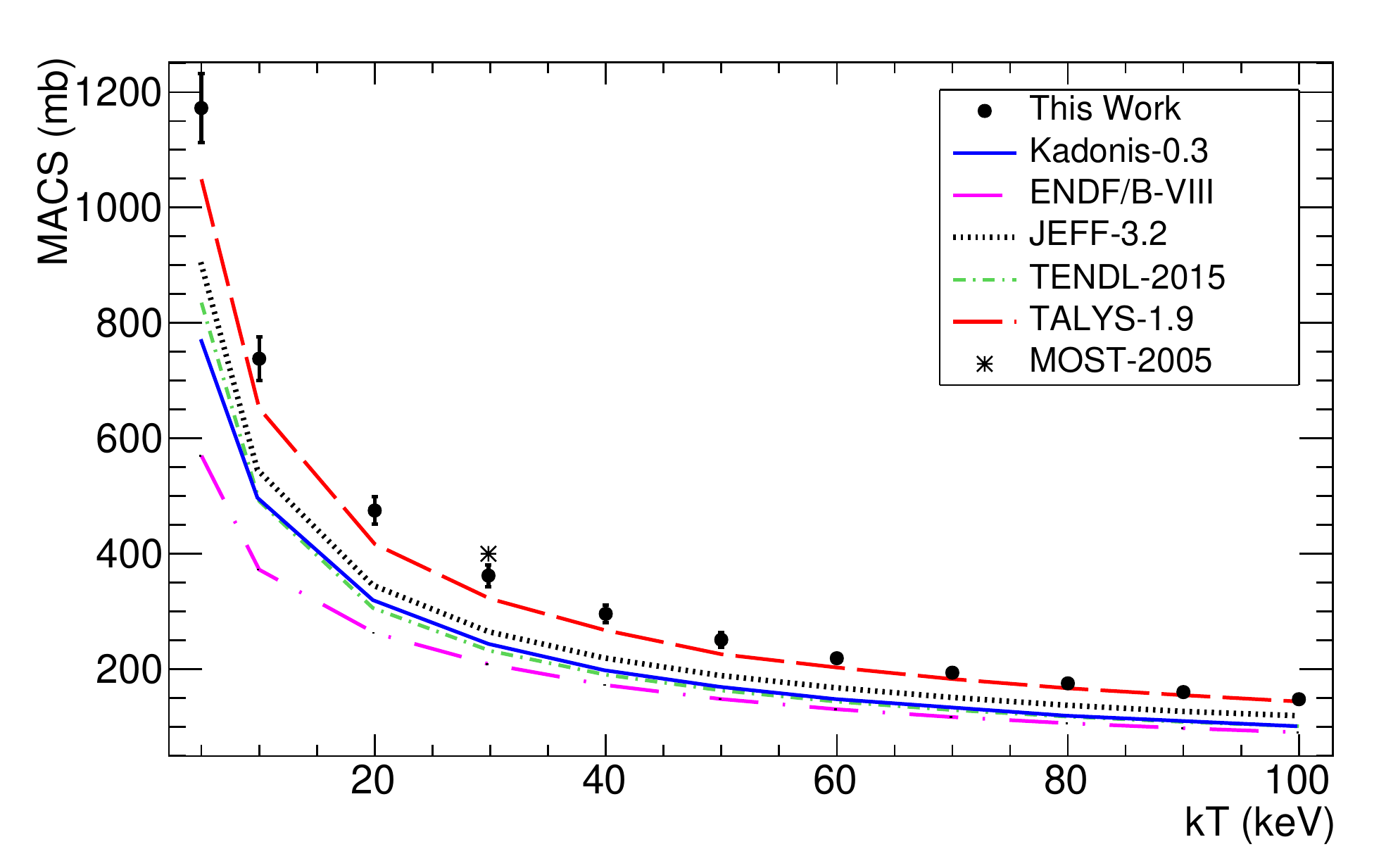}
\caption{Comparison of the experimental MACS from $kT=5-100$~keV with evaluations and theoretical predictions \cite{ENDFVIII,GOR05,TALYS,TENDL,JENDL,JEFF,kadonis}. 
MOST-2005 \cite{GOR05} and TALYS-1.9 \cite{TALYS} most closely reproduce the experimental
values, while all others significantly underestimate the MACS by a factor of up to 2 \label{macse}.}
\end{figure}

Above approximately 14 keV neutron energy, the experimental resolution became too low to resolve individual resonances. 
An averaged cross section was
determined from 14 keV to 300 keV neutron energy and self-shielding and
multiple scattering corrections were determined in Monte Carlo simulations.
These simulations followed the approach of tracing neutrons of a given energy
through the sample composed according to the specifications. The trace ended
either because the neutron got captured or left the sample. The energy loss
in each scattering step was considered, however purely isotropical scattering in
the center-of-mass frame was assumed. To determine the correction factors, we used neutron capture and scattering cross sections from ENDF-VIII \cite{ENDFVIII}, but scaled
the capture cross section to better match the experimentally determined cross sections. 
Corrections to the capture yield were always smaller than 6\%. Calculations 
using scaled and unscaled capture cross sections, different cross section evaluations, and changes in the sample thickness and geometry 
indicate that the total systematic uncertainty of the simulations is below 20\%, which results in at most 1.2\% uncertainty in the capture yield. \\
The cross sections reconstructed from SAMMY fits in the resolved resonance region below 14~keV, and the unresolved cross sections from 14 to 300~keV are shown in Fig. \ref{xsplots}.
Combining these two components, we calculated Maxwellian averaged cross sections (MACS) from $kT=5$ to $kT=100$~keV using
\begin{equation}
 MACS = \frac{2}{\sqrt{\pi}} \frac{1}{(kT)^{2}}\cdot \int_0^\infty E \sigma(E) \cdot \exp{\left(-\frac{E}{kT}\right)} \mbox{d}E
\end{equation}
For neutron energies above $300$ keV we used the evaluated ENDF-BV.III cross section \cite{ENDFVIII}, scaled by a factor of 1.7 to reproduce the experimental cross section 
at lower energies. The contribution of this energy range was at most 6\% (at $kT=100$~keV) and negligible for $kT<60$~keV. 
Fig. \ref{macse} shows a comparison of the experimental MACS from $kT=5-100$~keV, compared to recent evaluations and theoretical predictions. Besides TALYS-1.9 \cite{TALYS} (using default parameters), 
and MOST-2005 \cite{GOR05} the MACSs are significantly  underestimated by all predictions over the entire range of $kT$ values.  MACSs recommended by the Kadonis-0.3 database \cite{kadonis}, 
which is used in most nucleosynthesis calculations, are consistently a factor of about 1.5 lower over the entire energy range. 
MACSs from $kT=5$ to $kT=100$~keV are listed in Table \ref{macs}. Table \ref{uncert} summarises all contributions to the total uncertainty of these stellar cross sections. \\
In a star, excited states in nuclei may be thermally populated which means that the stellar reaction rate includes neutron capture on both, the ground state and excited states \cite{RAU12}. 
The cross sections on excited states have been estimated using TALYS-1.9 \cite{TALYS}, 
renormalising average level spacings and average radiative widths to the experimental values obtained in this work (see \ref{kerneltables}), and using the default 
optical model potential (OMP) \cite{KD03}. 
In addition, we investigated the impact on calculated cross sections using a different OMP, i.e. the JKM potential as described in Ref. \cite{BDG01}. 
Based on this we estimate a  factor 1.25 uncertainty 
for the theoretical neutron capture cross sections on excited states. Consequently, the stellar MACS$^{*}$, taking into account neutron capture on excited states, has uncertainties ranging from 
6\% at $kT=5$~keV, to 14\% at $kT=100$~keV. The MACS$^{*}$s are also listed in Table \ref{macs}.

\begin{table}[!htb]
\caption{\label{macs}Ground state Maxwellian Averaged Cross Sections on $^{73}$Ge, and stellar Maxwellian Averaged Cross Sections, taking into account 
neutron capture on thermally populated excited states. 
}
\begin{tabular}{|c|c|c|}
\hline
$kT$ (keV) & MACS (mb) & MACS$^{*}$ (mb) \\ \hline
5	& $	1170	\pm	60	$ & $	1174	\pm	69	$ \\
10	& $	738	\pm	38	$ & $	741	\pm	59	$ \\
20	& $	475	\pm	24	$ & $	470	\pm	48	$ \\
30	& $	362	\pm	19	$ & $	350	\pm	40	$ \\
40	& $	296	\pm	15	$ & $	281	\pm	34	$ \\
50	& $	251	\pm	13	$ & $	235	\pm	30	$ \\
60	& $	219	\pm	11	$ & $	203	\pm	27	$ \\
70	& $	194	\pm	10	$ & $	180	\pm	25	$ \\
80	& $	175.5	\pm	8.9	$ & $	162	\pm	23	$ \\
90	& $	160.4	\pm	8.2	$ & $	148	\pm	21	$ \\
100	& $	148.0	\pm	7.6	$ & $	136	\pm	20	$ \\
\hline
\end{tabular}

\end{table}

\begin{table}[!htb]
\caption{\label{uncert}Uncertainties of Maxwellian averaged $^{73}$Ge cross sections}
\begin{tabular}{cc}

Source & Uncertainty ($\%$) \\ \hline
Neutron Flux Shape &   \\
($<3$keV; 3 keV-1MeV) & $<1;3.5$  \\
Weighting Functions & 3 \\
Normalisation to Au &	1 \\
Background Subtraction& 1 \\
Sample Enrichment & 1 \\
Multiple Scattering and &  \\
Self Shielding ($>14$~keV)& 1.2 \\
Statistics & 0.3 \\ \hline
Total & 5.1 \\

\end{tabular}

\end{table}

\section{Astrophysical Implications}
The MACSs obtained in this work are a factor of about 1.5 larger than MACSs recommended to be used in stellar models, which were based on theoretical and semi-empirical estimates \cite{kadonis}. 
We have studied the impact of the new stellar neutron capture rate on weak $s$ process nucleosynthesis using a 15 solar mass ($M_\odot$) star with a metallicity of $Z=0.006$ (solar metallicity is $Z=0.014$ \cite{AGS09}), representative for 
a site with large overproduction of elemental germanium \cite{RHJ17}. In addition we have tested the new rates on main $s$ process nucleosynthesis in a 2 $M_\odot$ AGB star (Z=0.006). 
These calculations were performed using the multi-zone post processing code {\sc mppnp} \cite{HDF08}. 
It is estimated that the bulk of germanium in the solar system is produced in massive stars, while a small contribution of about 10-20\% comes from 
the main $s$ process in AGB stars \cite{PIG10}.
\begin{figure}[!htb]
\includegraphics[width=10.0 cm]{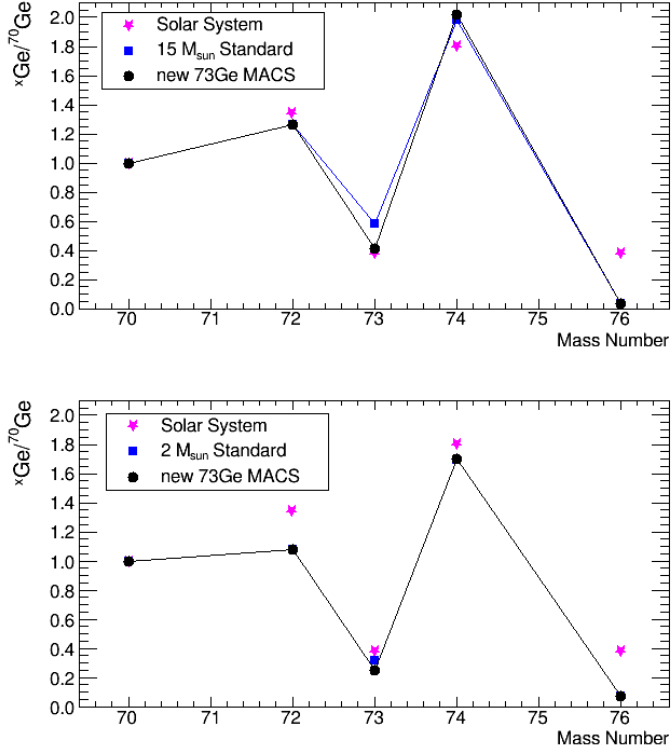}
\caption{(Top) Isotopic ratios rel. to $^{70}$Ge produced in a 15 $M_\odot$ star prior to Supernova explosion compared to solar system abundances (pink stars). Blue squares show the standard case and are compared
to results using the new $^{73}$Ge MACS (black circles). The new $^{73}$Ge MACS results in a lower $^{73}$Ge/$^{70}$Ge ratio,
consistent with the solar system isotopic ratio. (Bottom) Isotopic ratios produced in a 2 $M_\odot$ AGB star.  \label{gemass}}
\end{figure}
The top panel in Figure \ref{gemass} shows the isotopic abundance pattern of germanium isotopes produced in a 15 solar mass star prior to the supernova explosion. 
The germanium abundances are normalised to $^{70}$Ge in each case\footnote{$^{70}$Ge is mainly produced in the $s$ process 
and is shielded from rapid neutron  capture nucleosynthesis by stable $^{70}$Zn.}. The new $^{73}$Ge($n,\gamma$) MACS causes a reduction of  the $^{73}$Ge abundance by over
30\%, thus now reproducing the solar $^{73}$Ge/$^{70}$Ge ratio. $^{72}$Ge/$^{70}$Ge and $^{74}$Ge/$^{70}$Ge are reasonably close to the solar value, considering uncertainties in the associated
reaction rates and the fact that other nucleosynthesis processes contribute to a small extent to overall germanium abundances. $^{76}$Ge is significantly underproduced compared to solar 
as this isotope is bypassed by the $s$ process due to the short half life of $^{75}$Ge (terrestrial $t_{1/2}=83$ min) and is 
thought to be produced by explosive nucleosynthesis processes (see e.g. \cite{PRC14}). The bottom panel shows the same comparison for a  2$M_\odot$ AGB star. 
In this case, isotopic ratios are always smaller than solar.
Considering that the contribution of AGB stars to solar germanium is only about 10-20\%, we can expect that a combination of these two sites would still result in a fair reproduction
of the solar germanium abundance pattern.  To put firmer constraints on abundances produced in the different stellar sites, high precision MACS data on $^{70,72,74,76}$Ge are required; currently, there are no
neutron capture data on $^{72}$Ge above 4 keV, while uncertainties for recommended MACSs on $^{74,76}$Ge at $kT=30$~keV are 8-10\% \cite{kadonis1, MDD09}.  We expect that new n\_TOF data on these isotopes will 
improve the precision of these MACSs.
 \newline
 
  \section*{Acknowledgements}
This work was supported by the Austrian Science Fund FWF (J3503), the Adolf Messer Foundation (Germany), the UK Science and Facilities Council (ST/M006085/1), and the European Research Council ERC-2015-StG Nr. 677497. We acknowledge also support from MSMT of the Czech Republic.




\begin{thebibliography}{00}

\bibitem{PET68}
J.~G.~Peters, Astroph. J. {\bf 154}, 225 (1968).
\bibitem{COUCH74}
R.~G.~Couch, A.~B.~Schmiedekamp and W.~D.~Arnett, Astroph. J. {\bf 190}, 95 (1974).
\bibitem{LAMB77}
S.~A.~Lamb, W.~M.~Howard, J.~W.~Truran, and I.~Iben, Astroph. J. {\bf 217}, 213 (1977).
\bibitem{RAIT91a}
C.~M.~Raiteri, M.~Busso, G.~Picchio, and R.~Gallino, Astroph. J. {\bf 371}, 665 (1991).
\bibitem{RAIT91b}
C.~M.~Raiteri, M.~Busso, G.~Picchio, R.~Gallino, and L.~Pulone, Astroph. J. {\bf 367}, 228 (1991).

\bibitem{PIG10}
M.~Pignatari, R.~Gallino, M.~Heil, M.~Wiescher, F.~K\"appeler, F.~Herwig, S.~Bisterzo, Astroph. J. {\bf 710}, 1557-1577 (2010).
\bibitem{Nishi17} 
N. Nishimura, {\it et al.}, MNRAS {\bf 469}, 1752 (2017).

\bibitem{Mal68}
K.~Maletski, L.B.~Pikelner, I.M.~Salamatin, E.I. Sharapov, At. Energ. USSR {\bf 24},173 (1968).
\bibitem{HH80}
J.A. Harvey, M. Hockaday, EXFOR Entry 13770.004.
\bibitem{ENDFVIII}
D.A. Brown, {\it et al.}, Nucl. Data Sheets {\bf 148}, 1-142 (2018).




\bibitem{guerrero2012}
C. Guerrero, and {the n\_TOF collaboration}, Eur. Phys. J. A {\bf48}, 29(2012).

\bibitem{MBK11}
C. Massimi, {\it et al.},  J. Kor. Phys. Soc. {\bf 59}, 1689 (2011).

\bibitem{pwht1}
R.L. Macklin and R.H. Gibbons, Phys. Rev. {\bf 159}, 1007 (1967).

\bibitem{pwht2}
U. Abbondanno, and the n\_TOF collaboration, Nucl. Instrum. Methods Phys. Res. A {\bf 521}, 454–467 (2004).

\bibitem{geant4}
S. Agostinelli {\it et~al.} (Geant4 Collaboration), Nucl. Instr. Meth. Phys. Res. {\bf A 506}, 250 (2003).

\bibitem{Bec98}
F.~Be\v{c}v\'{a}\v{r}, Nucl. Instr. Meth. A{\bf 417}, 434 (1998).


\bibitem{BS16}
M.~Sabat\'{e}-Gilarte, M. Barbagallo, and the n\_TOF Collaboration. n\_TOF Collaboration Meeting, Catania, May 2016. 

\bibitem{Barb13}
M. Barbagallo, {\it et al.}, Eur. Phys. J. A {\bf 49}, 156 (2013).


\bibitem{sammy}
N.M. Larson, Technical report ORNL/TM-9179/R8, {\it Updated users guide for SAMMY: Multilevel R-matrix 
	 fits to neutron data using Bayes' equations}, Oak-Ridge National Laboratory, Oak Ridge, TN, USA (2008).
	 


\bibitem{GOR05}
S. Goriely, Hauser-Feshbach rates for neutron capture reactions (version 8/29/2005), http://www-astro.ulb.ac.be/Html/hfr.html. 

\bibitem{TALYS}
A.J. Koning, {\it et al.}, TALYS-1.9, online at  www.talys.eu.

\bibitem{TENDL}
A.J. Koning, {\it et al.}, Nucl. Data Sheets {\bf 113}, 2841 (2012) (https://tendl.web.psi.ch/tendl\_2015/tendl2015.html).

\bibitem{JENDL}
K. Shibata, {\it et al.}, J. Nucl. Sci. Technol.  {\bf 48}, 1 (2011).

\bibitem{JEFF}
JEFF-3.2 (2014) available online: http://www.oecd-nea.org/dbforms/data/eva/evatapes/jeff\_32/.



\bibitem{RAU12}
T.~Rauscher, Astroph. J. Letters {\bf 755}, L10 (2012); Astroph. J. Letters {\bf 864}, L40 (2018).

\bibitem{KD03}
A.J. Koning and J.P. Delaroche, Nucl. Phys. A {\bf 713}, 231 (2003).

\bibitem{BDG01}
E. Bauge, J.P Delaroche, and M. Girod, Phys. Rev. C{\bf 63}, 024607 (2001).

\bibitem{kadonis}
I.~Dillmann, R.~Plag, F.~K{\"a}ppeler, and T.~Rauscher, {\it EFNUDAT Fast Neutrons - scientific workshop on neutron measurements, theory \& applications,} JRC-IRMM (2009); online at http://www.kadonis.org.


\bibitem{AGS09}
M. Asplund, {\it et al.}, Ann. Rev. Astron. Astrophys.   {\bf 47}, 481 (2009).

\bibitem{RHJ17}
C. Ritter,  {\it et al.}, MNRAS {\bf 480}, 538 (2018).

\bibitem{HDF08}
F. Herwig, {\it et al.}, PoS(NIC X)023 (2008).  

\bibitem{PRC14}
A. Perego,  {\it et al.}, MNRAS   {\bf 443}, 3134 (2014).

\bibitem{kadonis1}
KADoNis Database Revision 1.0 (test version).

\bibitem{MDD09}
J. Marganiec, {\it et al.}, Phys. Rev. C{\bf 79}, 065802 (2009).


\bibitem{Capote09} R. Capote, {\it et al}, Nucl. Data Sheets {\bf 110}, 3107 (2009).
\bibitem{Mughabghab} S. Mughabghab, Atlas of Neutron Resonances, 5th Edition, Elsevier, 2006.
\bibitem{Lerendegui-Marco18} J. Lerendegui-Marco, {\it et al}, Phys. Rev. C{\bf 97}, 024605 (2018).







\end{thebibliography}

\appendix

\section{Resonance Data \label{kerneltables}} 
Tables \ref{kernels} to \ref{kernels5} list capture kernels $k$ and associated fit uncertainties. The resonance energies have a further
systematic uncertainty of $0.06\%$ due to the uncertainty in the neutron flight path. The systematic uncertainty on the capture kernels are 3.5\% below and 4.8\% above 3~keV neutron energy,
consisting of uncertainties due to Pulse Height Weighting (3\%), normalisation (1\%), sample enrichment (1\%), and neutron flux (1\% below, 3.5\% above 3~keV).
Neutron capture data by themselves do not always allow a reliable determination of neutron and radiative widths and resonance spins. 
However, the results from SAMMY fits were consistent with $\langle \Gamma_\gamma \rangle \approx 250$ meV 
for all resonances with $E_R<8$ keV and $\Gamma_n/\Gamma_\gamma>>1$; specifically, the maximum-likelihood 
estimate assuming a Gaussian distribution of radiative widths yielded $\langle \Gamma_\gamma \rangle=250(10)$~meV and $\sigma_{\Gamma_\gamma}=30(5)$~meV. 
Statistical model simulations using the {\sc dicebox} code~\cite{Bec98} indicated only a slightly narrower distribution ($15-20$~meV). 
The present experimental value is somewhat higher than literature values of $\langle \Gamma_\gamma \rangle=195(45)$~meV~\cite{Capote09} and $197(6)$~meV~\cite{Mughabghab}.

\begin{figure}[!htb]
\includegraphics[width=8.0 cm]{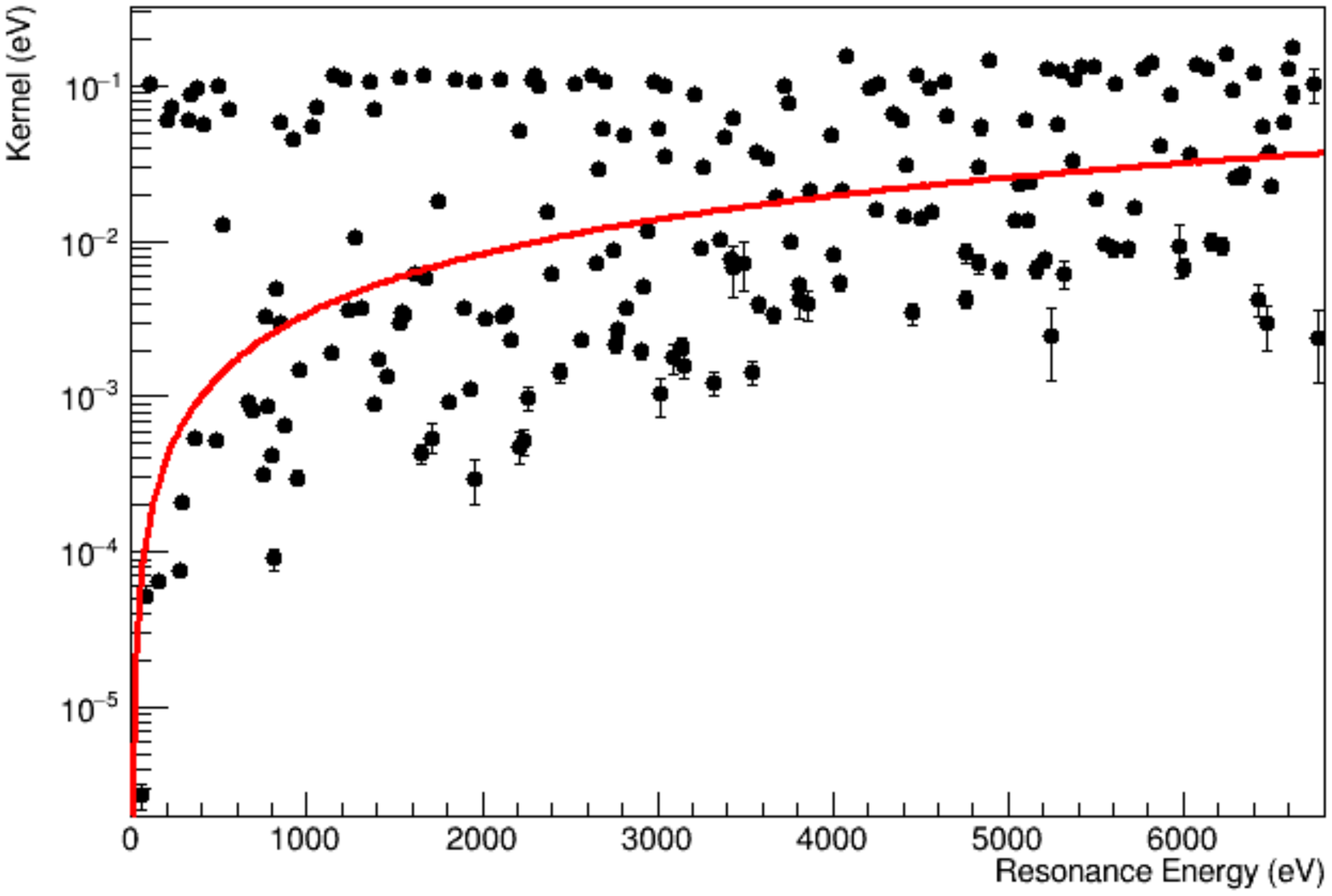}
\caption{Capture kernels $k$ of $^{73}$Ge($n,\gamma$) resonances. The solid line shows  the bias function (equiprobability for a single resonance at a given neutron energy to be either
$s$- or $p$-wave), constructed to deduce statistical resonance properties. \label{kernelplot}}
\end{figure}

To deduce the statistical resonance properties we construct the so-called bias function, that is the equiprobability for a single resonance at a given neutron energy to be either
$s$- or $p$-wave. The bias function, as shown in Fig.~\ref{kernelplot}, is obtained using the present value of $\langle\Gamma_\gamma\rangle$ and varying values of the
resonance spacing $D_0$, the $s$-wave  neutron strength function $S_0$  and the $p$-wave  neutron strength function $S_1$. The channel radius was taken in the form
$R=1.35A^{1/3}$.  The comparison of the number of resonances with $k$  above different multiples ($1-5\times$) of the bias function yielded $D_0=70(8)$~eV,
$S_0=1.90(25)\times10^{-4}$  and $S_1=1.1(3)\times10^{-4}$;  the latter value scales with $R$  as $R^2\times S_1 =$ const. For comparison, literature values are
$D_0=62(15)$  eV, $S_0=2.0(4)\times10^{-4}$ from Ref.~\cite{Capote09}, and $D_0=99(10)$  eV, $S_0=1.66(40)\times 10^{-4}$  from Ref.~\cite{Mughabghab}.
 Present values were obtained similarly to Ref.~\cite{Lerendegui-Marco18} from comparison of experimental  data with simulations of individual resonance sequences 
 using Wigner distribution of level spacings, Porter-Thomas distribution of reduced neutron widths, and Gaussian distribution of radiative widths.

\begin{table}[!htb]
\caption{\label{kernels}Resonance energies and capture kernels $k$ of  $^{73}$Ge($n,\gamma$) $^{a}$Resonances fitted with natural germanium sample. $^{*}$Resonances listed in ENDF/B-VIII \cite{ENDFVIII}. 
$^{b}$Doublet in ENDF/B-VIII \cite{ENDFVIII}.
}
\begin{tabular}{|c|c|c|c|c|c|}
\hline
$	E_R$ (eV)				&	$	k$ (meV)				&		&	$	E_R$ (eV)	&	$	k$ (meV)	\\ \hline
$	59.34	\pm	0.09	$	&	$	0.0027	\pm	0.0005	$	&		&	$	1219.98	\pm	0.02	^{*}$	&	$	110.01	\pm	0.98	$	\\
$	79.47	\pm	0.02	$	&	$	0.051	\pm	0.002	$	&		&	$	1233.15	\pm	0.05	$	&	$	10.56	\pm	0.39	$	\\
$	102.73	\pm	0.01	^{a,*}$	&	$	104.56	\pm	0.63	$	&		&	$	1276.91	\pm	0.05	$	&	$	3.70	\pm	0.13	$	\\
$	156.32	\pm	0.02	$	&	$	0.064	\pm	0.003	$	&		&	$	1316.76	\pm	0.03	^{*}$	&	$	107.28	\pm	0.92	$	\\
$	204.17	\pm	0.01	^{a,*}$	&	$	60.87	\pm	0.54	$	&		&	$	1358.03	\pm	0.02	^{*}$	&	$	70.77	\pm	0.78	$	\\
$	224.83	\pm	0.01	^{*}$	&	$	71.82	\pm	0.21	$	&		&	$	1380.23	\pm	0.19	$	&	$	0.90	\pm	0.10	$	\\
$	270.90	\pm	0.06	$	&	$	0.075	\pm	0.006	$	&		&	$	1384.68	\pm	0.10	$	&	$	1.73	\pm	0.11	$	\\
$	286.68	\pm	0.02	$	&	$	0.206	\pm	0.007	$	&		&	$	1404.78	\pm	0.10	$	&	$	1.36	\pm	0.09	$	\\
$	320.70	\pm	0.01	^{a,*}$	&	$	59.45	\pm	0.92	$	&		&	$	1462.39	\pm	0.08	$	&	$	2.97	\pm	0.15	$	\\
$	332.85	\pm	0.02	^{a,*}$	&	$	88.05	\pm	1.39	$	&		&	$	1530.62	\pm	0.09	^{*}$	&	$	115.19	\pm	3.57	$	\\
$	361.82	\pm	0.06	$	&	$	0.533	\pm	0.046	$	&		&	$	1540.42	\pm	0.10	$	&	$	3.44	\pm	0.22	$	\\
$	368.08	\pm	0.01	^{*}$	&	$	97.68	\pm	0.41	$	&		&	$	1549.31	\pm	0.14	$	&	$	3.42	\pm	0.33	$	\\
$	409.28	\pm	0.01	^{*}$	&	$	56.75	\pm	0.25	$	&		&	$	1552.26	\pm	0.07	$	&	$	6.19	\pm	0.24	$	\\
$	479.55	\pm	0.06	$	&	$	0.52	\pm	0.03	$	&		&	$	1614.59	\pm	0.26	$	&	$	0.43	\pm	0.06	$	\\
$	491.55	\pm	0.02	^{*}$	&	$	101.31	\pm	0.54	$	&		&	$	1655.95	\pm	0.05	^{*}$	&	$	118.38	\pm	1.26	$	\\
$	517.99	\pm	0.01	^{*}$	&	$	12.83	\pm	0.13	$	&		&	$	1665.35	\pm	0.09	$	&	$	5.87	\pm	0.31	$	\\
$	558.34	\pm	0.01	^{*}$	&	$	71.62	\pm	0.32	$	&		&	$	1675.01	\pm	0.45	$	&	$	0.55	\pm	0.12	$	\\
$	668.80	\pm	0.04	^{*}$	&	$	0.92	\pm	0.03	$	&		&	$	1716.18	\pm	0.04	$	&	$	18.26	\pm	0.38	$	\\
$	693.58	\pm	0.04	$	&	$	0.82	\pm	0.03	$	&		&	$	1751.88	\pm	0.19	$	&	$	0.93	\pm	0.09	$	\\
$	750.31	\pm	0.08	$	&	$	0.31	\pm	0.02	$	&		&	$	1808.02	\pm	0.04	^{*}$	&	$	109.75	\pm	1.15	$	\\
$	762.70	\pm	0.02	$	&	$	3.25	\pm	0.06	$	&		&	$	1843.00	\pm	0.08	$	&	$	3.71	\pm	0.18	$	\\
$	777.48	\pm	0.04	$	&	$	0.87	\pm	0.03	$	&		&	$	1897.71	\pm	0.18	$	&	$	1.12	\pm	0.10	$	\\
$	798.47	\pm	0.06	$	&	$	0.42	\pm	0.03	$	&		&	$	1930.82	\pm	0.74	$	&	$	0.30	\pm	0.09	$	\\
$	816.73	\pm	0.22	$	&	$	0.09	\pm	0.01	$	&		&	$	1952.79	\pm	0.05	^{b,*}$	&	$	105.78	\pm	1.27	$	\\
$	826.23	\pm	0.02	$	&	$	5.00	\pm	0.09	$	&		&	$	1963.43	\pm	0.15	$	&	$	3.21	\pm	0.29	$	\\
$	843.31	\pm	0.03	$	&	$	2.96	\pm	0.09	$	&		&	$	2019.34	\pm	0.06	^{*}$	&	$	111.08	\pm	1.55	$	\\
$	851.39	\pm	0.01	^{*}$	&	$	57.96	\pm	0.47	$	&		&	$	2104.02	\pm	0.11	$	&	$	3.28	\pm	0.20	$	\\
$	878.76	\pm	0.08	$	&	$	0.64	\pm	0.04	$	&		&	$	2116.60	\pm	0.13	$	&	$	3.49	\pm	0.21	$	\\
$	920.46	\pm	0.01	^{*}$	&	$	45.96	\pm	0.43	$	&		&	$	2144.13	\pm	0.17	$	&	$	2.34	\pm	0.18	$	\\
$	948.07	\pm	0.13	$	&	$	0.30	\pm	0.03	$	&		&	$	2162.03	\pm	0.04	$	&	$	51.19	\pm	0.99	$	\\
$	959.17	\pm	0.05	$	&	$	1.49	\pm	0.06	$	&		&	$	2211.39	\pm	0.59	$	&	$	0.48	\pm	0.11	$	\\
$	1031.14	\pm	0.01	^{*}$	&	$	55.02	\pm	0.59	$	&		&	$	2216.97	\pm	0.51	$	&	$	0.51	\pm	0.10	$	\\
$	1059.02	\pm	0.01	$	&	$	72.65	\pm	0.67	$	&		&	$	2236.71	\pm	0.42	$	&	$	0.99	\pm	0.17	$	\\
$	1139.18	\pm	0.12	$	&	$	1.93	\pm	0.16	$	&		&	$	2262.54	\pm	0.06	^{*}$	&	$	109.62	\pm	1.73	$	\\
$	1148.29	\pm	0.03	^{*}$	&	$	116.35	\pm	0.93	$	&		&	$	2291.05	\pm	0.13	$	&	$	117.25	\pm	4.30	$	\\ \hline

\end{tabular}

\end{table}

\newpage
\begin{table}[!htb]
\caption{Table \ref{kernels} continued.
\label{kernels2}}
\begin{tabular}{|c|c|c|c|c|c|}
\hline
$	E_R$ (eV)				&	$	k$ (meV)				&		&	$	E_R$ (eV)	&	$	k$ (meV)	\\ \hline
$	2297.90	\pm	0.07	^{*}$	&	$	98.62	\pm	4.11	$	&		&	$	3431.09	\pm	0.66	$	&	$	6.80	\pm	2.47	$	\\
$	2321.94	\pm	0.07	$	&	$	15.32	\pm	0.52	$	&		&	$	3432.30	\pm	0.59	$	&	$	7.24	\pm	2.53	$	\\
$	2373.25	\pm	0.12	$	&	$	6.25	\pm	0.32	$	&		&	$	3495.70	\pm	0.57	$	&	$	1.45	\pm	0.25	$	\\
$	2398.90	\pm	0.30	$	&	$	1.42	\pm	0.19	$	&		&	$	3535.63	\pm	0.08	$	&	$	36.88	\pm	1.03	$	\\
$	2442.65	\pm	0.04	$	&	$	102.11	\pm	1.52	$	&		&	$	3565.08	\pm	0.34	$	&	$	3.90	\pm	0.41	$	\\
$	2531.55	\pm	0.24	$	&	$	2.30	\pm	0.23	$	&		&	$	3579.46	\pm	0.08	$	&	$	34.21	\pm	1.05	$	\\
$	2566.67	\pm	0.06	^{*}$	&	$	118.78	\pm	1.74	$	&		&	$	3628.35	\pm	0.34	$	&	$	3.36	\pm	0.36	$	\\
$	2624.55	\pm	0.14	$	&	$	7.24	\pm	0.44	$	&		&	$	3657.54	\pm	0.16	$	&	$	19.55	\pm	0.92	$	\\
$	2648.04	\pm	0.06	^{*}$	&	$	29.34	\pm	0.85	$	&		&	$	3674.52	\pm	0.07	$	&	$	99.37	\pm	2.30	$	\\
$	2666.79	\pm	0.05	$	&	$	53.87	\pm	1.37	$	&		&	$	3718.11	\pm	0.07	$	&	$	77.49	\pm	1.95	$	\\
$	2688.08	\pm	0.05	$	&	$	107.17	\pm	1.92	$	&		&	$	3745.16	\pm	0.19	$	&	$	10.03	\pm	0.62	$	\\
$	2697.89	\pm	0.14	$	&	$	8.76	\pm	0.57	$	&		&	$	3763.26	\pm	0.25	$	&	$	5.31	\pm	0.44	$	\\
$	2748.81	\pm	0.40	^{*}$	&	$	2.20	\pm	0.30	$	&		&	$	3809.15	\pm	0.59	$	&	$	4.15	\pm	0.97	$	\\
$	2763.67	\pm	0.29	$	&	$	2.71	\pm	0.31	$	&		&	$	3812.80	\pm	0.46	$	&	$	3.94	\pm	0.85	$	\\
$	2776.01	\pm	0.06	$	&	$	47.66	\pm	1.08	$	&		&	$	3852.87	\pm	0.15	$	&	$	20.94	\pm	0.94	$	\\
$	2806.78	\pm	0.18	$	&	$	3.77	\pm	0.29	$	&		&	$	3870.01	\pm	0.09	$	&	$	47.98	\pm	1.40	$	\\
$	2821.52	\pm	0.36	$	&	$	1.98	\pm	0.22	$	&		&	$	3993.96	\pm	0.29	$	&	$	8.22	\pm	0.71	$	\\
$	2903.85	\pm	0.18	$	&	$	5.03	\pm	0.35	$	&		&	$	4002.66	\pm	0.16	$	&	$	5.44	\pm	0.58	$	\\
$	2924.80	\pm	0.14	$	&	$	11.56	\pm	0.61	$	&		&	$	4042.61	\pm	0.26	$	&	$	21.00	\pm	2.02	$	\\
$	2946.18	\pm	0.09	$	&	$	107.59	\pm	1.97	$	&		&	$	4053.83	\pm	0.25	$	&	$	155.04	\pm	4.40	$	\\
$	2982.63	\pm	0.06	$	&	$	52.53	\pm	1.24	$	&		&	$	4073.85	\pm	0.09	$	&	$	95.40	\pm	3.39	$	\\
$	3005.31	\pm	0.85	^{*}$	&	$	1.03	\pm	0.29	$	&		&	$	4215.88	\pm	0.20	$	&	$	16.14	\pm	0.86	$	\\
$	3023.45	\pm	0.05	$	&	$	99.68	\pm	1.95	$	&		&	$	4246.57	\pm	0.25	$	&	$	103.52	\pm	4.11	$	\\
$	3037.63	\pm	0.08	$	&	$	35.20	\pm	1.18	$	&		&	$	4254.81	\pm	0.12	$	&	$	67.18	\pm	3.49	$	\\
$	3044.63	\pm	0.63	$	&	$	1.78	\pm	0.40	$	&		&	$	4349.23	\pm	0.10	^{*}$	&	$	61.07	\pm	1.78	$	\\
$	3085.74	\pm	0.37	$	&	$	2.11	\pm	0.26	$	&		&	$	4394.74	\pm	0.22	$	&	$	14.48	\pm	0.93	$	\\
$	3133.36	\pm	0.57	$	&	$	1.60	\pm	0.29	$	&		&	$	4406.67	\pm	0.14	$	&	$	30.98	\pm	1.43	$	\\
$	3155.23	\pm	0.06	$	&	$	88.69	\pm	1.65	$	&		&	$	4419.23	\pm	0.57	^{*}$	&	$	3.44	\pm	0.53	$	\\
$	3214.07	\pm	0.15	$	&	$	8.96	\pm	0.52	$	&		&	$	4458.61	\pm	0.13	$	&	$	118.69	\pm	2.81	$	\\
$	3251.24	\pm	0.07	$	&	$	30.04	\pm	0.90	$	&		&	$	4475.90	\pm	0.28	$	&	$	14.18	\pm	1.01	$	\\
$	3264.48	\pm	0.43	$	&	$	1.21	\pm	0.20	$	&		&	$	4499.19	\pm	0.11	$	&	$	98.06	\pm	2.58	$	\\
$	3320.71	\pm	0.17	$	&	$	10.34	\pm	0.63	$	&		&	$	4548.89	\pm	0.26	$	&	$	15.23	\pm	1.09	$	\\
$	3361.67	\pm	0.10	$	&	$	46.80	\pm	1.48	$	&		&	$	4564.61	\pm	0.10	$	&	$	105.98	\pm	2.66	$	\\
$	3388.19	\pm	0.24	$	&	$	7.64	\pm	0.56	$	&		&	$	4635.76	\pm	0.10	$	&	$	64.60	\pm	2.00	$	\\
$	3414.94	\pm	0.21	$	&	$	61.45	\pm	4.57	$	&		&	$	4649.59	\pm	0.46	$	&	$	4.28	\pm	0.54	$	\\
 \hline
\end{tabular}

\end{table}

\newpage
\begin{table}[!htb]
\caption{Table \ref{kernels} continued.
\label{kernels3}}
\begin{tabular}{|c|c|c|c|c|c|}
\hline
$	E_R$ (eV)				&	$	k$ (meV)				&		&	$	E_R$ (eV)	&	$	k$ (meV)	\\ \hline
$	4753.89	\pm	0.47	$	&	$	8.45	\pm	1.21	$	&		&	$	6040.28	\pm	0.15	$	&	$	136.59	\pm	4.32	$	\\
$	4759.64	\pm	0.47	$	&	$	7.25	\pm	1.07	$	&		&	$	6075.56	\pm	0.16	^{*}$	&	$	129.63	\pm	4.05	$	\\
$	4827.94	\pm	0.29	$	&	$	29.67	\pm	1.96	$	&		&	$	6136.50	\pm	0.65	$	&	$	9.92	\pm	1.18	$	\\
$	4837.08	\pm	0.35	$	&	$	54.47	\pm	5.93	$	&		&	$	6155.03	\pm	0.59	$	&	$	9.35	\pm	1.15	$	\\
$	4843.10	\pm	0.17	$	&	$	145.17	\pm	7.29	$	&		&	$	6219.65	\pm	0.17	$	&	$	162.57	\pm	4.94	$	\\
$	4898.08	\pm	0.50	$	&	$	6.53	\pm	0.75	$	&		&	$	6241.35	\pm	0.16	$	&	$	92.48	\pm	3.85	$	\\
$	4949.81	\pm	0.29	$	&	$	13.67	\pm	1.02	$	&		&	$	6279.25	\pm	0.41	$	&	$	25.25	\pm	2.20	$	\\
$	5039.54	\pm	0.22	$	&	$	23.53	\pm	1.29	$	&		&	$	6289.68	\pm	0.39	$	&	$	25.43	\pm	2.26	$	\\
$	5061.31	\pm	0.14	$	&	$	60.97	\pm	2.29	$	&		&	$	6327.02	\pm	0.41	$	&	$	27.42	\pm	2.22	$	\\
$	5098.11	\pm	0.41	$	&	$	13.75	\pm	1.31	$	&		&	$	6344.27	\pm	0.19	$	&	$	119.87	\pm	4.03	$	\\
$	5106.77	\pm	0.22	^{*}$	&	$	24.34	\pm	1.53	$	&		&	$	6400.37	\pm	1.31	$	&	$	4.25	\pm	1.00	$	\\
$	5126.34	\pm	0.44	$	&	$	6.58	\pm	0.77	$	&		&	$	6427.65	\pm	0.19	^{*}$	&	$	54.35	\pm	2.44	$	\\
$	5160.94	\pm	0.42	$	&	$	7.58	\pm	0.79	$	&		&	$	6455.88	\pm	1.73	$	&	$	2.93	\pm	0.94	$	\\
$	5207.94	\pm	0.17	$	&	$	130.66	\pm	3.75	$	&		&	$	6470.60	\pm	0.28	$	&	$	36.98	\pm	2.27	$	\\
$	5221.27	\pm	0.29	$	&	$	2.48	\pm	1.21	$	&		&	$	6485.66	\pm	0.39	$	&	$	22.75	\pm	1.86	$	\\
$	5245.99	\pm	0.14	$	&	$	56.01	\pm	2.37	$	&		&	$	6504.18	\pm	0.18	$	&	$	57.68	\pm	2.52	$	\\
$	5281.75	\pm	0.13	$	&	$	124.70	\pm	3.34	$	&		&	$	6578.25	\pm	0.25	$	&	$	128.75	\pm	5.10	$	\\
$	5302.86	\pm	0.89	$	&	$	6.23	\pm	1.30	$	&		&	$	6602.37	\pm	0.38	$	&	$	88.44	\pm	10.83	$	\\
$	5312.62	\pm	0.23	$	&	$	32.68	\pm	1.97	$	&		&	$	6617.24	\pm	0.32	$	&	$	179.31	\pm	7.07	$	\\
$	5364.07	\pm	0.24	$	&	$	110.95	\pm	4.84	$	&		&	$	6626.46	\pm	0.68	$	&	$	102.51	\pm	24.32	$	\\
$	5381.79	\pm	0.24	^{*}$	&	$	134.42	\pm	5.45	$	&		&	$	6741.61	\pm	0.52	$	&	$	2.41	\pm	1.20	$	\\
$	5419.94	\pm	0.12	$	&	$	134.51	\pm	3.65	$	&		&	$	6766.29	\pm	0.17	$	&	$	161.31	\pm	5.16	$	\\
$	5485.70	\pm	0.28	$	&	$	18.42	\pm	1.27	$	&		&	$	6826.24	\pm	0.85	$	&	$	8.84	\pm	1.19	$	\\
$	5506.14	\pm	0.38	$	&	$	9.76	\pm	0.89	$	&		&	$	6866.12	\pm	0.22	$	&	$	109.52	\pm	3.86	$	\\
$	5547.03	\pm	0.45	$	&	$	8.88	\pm	0.97	$	&		&	$	6883.56	\pm	0.60	$	&	$	15.78	\pm	4.46	$	\\
$	5598.25	\pm	0.14	^{*}$	&	$	103.02	\pm	6.29	$	&		&	$	6955.68	\pm	0.19	$	&	$	90.86	\pm	3.45	$	\\
$	5612.71	\pm	0.42	^{*}$	&	$	9.01	\pm	1.05	$	&		&	$	6973.99	\pm	0.61	$	&	$	11.48	\pm	1.51	$	\\
$	5681.92	\pm	0.37	$	&	$	16.51	\pm	1.29	$	&		&	$	7118.07	\pm	0.21	^{*}$	&	$	81.88	\pm	3.25	$	\\
$	5716.47	\pm	0.14	$	&	$	127.76	\pm	3.63	$	&		&	$	7154.02	\pm	0.15	$	&	$	12.70	\pm	4.38	$	\\
$	5764.60	\pm	0.23	$	&	$	142.28	\pm	4.41	$	&		&	$	7177.26	\pm	0.38	$	&	$	42.10	\pm	2.81	$	\\
$	5815.35	\pm	0.22	$	&	$	41.61	\pm	2.36	$	&		&	$	7221.40	\pm	0.18	$	&	$	127.16	\pm	4.61	$	\\
$	5873.42	\pm	0.18	$	&	$	88.73	\pm	3.26	$	&		&	$	7245.57	\pm	0.52	$	&	$	24.75	\pm	2.42	$	\\
$	5927.43	\pm	0.03	$	&	$	9.23	\pm	3.44	$	&		&	$	7290.68	\pm	0.27	$	&	$	138.57	\pm	5.11	$	\\
$	5975.46	\pm	0.65	$	&	$	6.73	\pm	0.85	$	&		&	$	7330.83	\pm	0.32	$	&	$	138.82	\pm	5.48	$	\\
$	6002.51	\pm	0.27	$	&	$	35.82	\pm	2.02	$	&		&	$	7370.59	\pm	0.24	$	&	$	90.99	\pm	4.04	$	\\
\hline

\end{tabular}

\end{table}

\newpage
\begin{table}[!htb]
\caption{Table \ref{kernels} continued.
\label{kernels4}}
\begin{tabular}{|c|c|c|c|c|c|}
\hline
$	E_R$ (eV)				&	$	k$ (meV)				&		&	$	E_R$ (eV)	&	$	k$ (meV)	\\ \hline
$	7480.92	\pm	0.20	$	&	$	71.03	\pm	3.10	$	&		&	$	9038.39	\pm	0.25	$	&	$	129.28	\pm	3.32	$	\\
$	7508.04	\pm	1.36	$	&	$	7.98	\pm	1.55	$	&		&	$	9086.33	\pm	0.55	$	&	$	50.07	\pm	4.45	$	\\
$	7601.45	\pm	0.23	$	&	$	88.12	\pm	3.48	$	&		&	$	9097.14	\pm	0.92	$	&	$	22.78	\pm	3.63	$	\\
$	7641.78	\pm	0.21	$	&	$	138.99	\pm	4.79	$	&		&	$	9147.22	\pm	1.65	$	&	$	16.07	\pm	2.80	$	\\
$	7756.02	\pm	0.81	$	&	$	13.71	\pm	1.80	$	&		&	$	9168.58	\pm	0.09	$	&	$	12.59	\pm	5.52	$	\\
$	7777.46	\pm	0.35	$	&	$	35.77	\pm	2.45	$	&		&	$	9191.30	\pm	1.04	$	&	$	26.99	\pm	3.82	$	\\
$	7830.29	\pm	0.63	$	&	$	24.88	\pm	2.35	$	&		&	$	9209.69	\pm	0.98	$	&	$	51.89	\pm	9.67	$	\\
$	7862.45	\pm	0.50	$	&	$	107.18	\pm	6.19	$	&		&	$	9222.47	\pm	0.57	$	&	$	238.54	\pm	15.55	$	\\
$	7898.08	\pm	0.23	$	&	$	152.34	\pm	6.44	$	&		&	$	9266.86	\pm	0.61	$	&	$	46.08	\pm	4.38	$	\\
$	7931.41	\pm	0.28	$	&	$	66.22	\pm	3.58	$	&		&	$	9319.08	\pm	0.48	$	&	$	266.48	\pm	10.47	$	\\
$	7971.27	\pm	0.31	$	&	$	132.45	\pm	5.25	$	&		&	$	9344.08	\pm	0.34	$	&	$	117.15	\pm	7.25	$	\\
$	8020.47	\pm	0.32	$	&	$	87.03	\pm	5.64	$	&		&	$	9363.08	\pm	0.06	$	&	$	16.84	\pm	6.85	$	\\
$	8036.76	\pm	0.52	$	&	$	29.75	\pm	2.78	$	&		&	$	9490.32	\pm	0.27	$	&	$	171.80	\pm	6.02	$	\\
$	8099.71	\pm	0.26	$	&	$	172.56	\pm	5.65	$	&		&	$	9545.52	\pm	1.19	$	&	$	16.24	\pm	3.16	$	\\
$	8148.02	\pm	0.46	$	&	$	38.81	\pm	2.86	$	&		&	$	9559.05	\pm	0.96	$	&	$	24.89	\pm	3.58	$	\\
$	8180.99	\pm	0.29	$	&	$	89.16	\pm	4.39	$	&		&	$	9575.23	\pm	0.83	$	&	$	25.93	\pm	3.44	$	\\
$	8329.05	\pm	0.37	$	&	$	41.98	\pm	2.77	$	&		&	$	9626.54	\pm	0.30	$	&	$	114.82	\pm	5.06	$	\\
$	8349.70	\pm	0.18	$	&	$	7.18	\pm	3.33	$	&		&	$	9707.30	\pm	1.76	$	&	$	10.11	\pm	3.10	$	\\
$	8417.95	\pm	0.11	^{*}$	&	$	3.14	\pm	1.58	$	&		&	$	9728.35	\pm	0.38	$	&	$	100.08	\pm	5.62	$	\\
$	8442.04	\pm	0.91	$	&	$	12.93	\pm	1.96	$	&		&	$	9802.54	\pm	0.85	$	&	$	105.66	\pm	9.33	$	\\
$	8488.63	\pm	2.31	$	&	$	84.68	\pm	17.73	$	&		&	$	9816.46	\pm	0.47	$	&	$	98.85	\pm	8.42	$	\\
$	8507.36	\pm	1.15	$	&	$	60.74	\pm	15.53	$	&		&	$	9863.34	\pm	0.74	$	&	$	22.36	\pm	2.80	$	\\
$	8543.29	\pm	0.54	$	&	$	25.55	\pm	2.63	$	&		&	$	9965.97	\pm	0.12	$	&	$	28.25	\pm	10.94	$	\\
$	8587.34	\pm	0.00	$	&	$	102.17	\pm	0.00	$	&		&	$	9994.54	\pm	0.74	$	&	$	76.06	\pm	8.59	$	\\
$	8608.28	\pm	0.11	$	&	$	27.93	\pm	10.68	$	&		&	$	10009.71	\pm	0.40	$	&	$	160.40	\pm	8.59	$	\\
$	8643.80	\pm	0.13	$	&	$	6.27	\pm	2.98	$	&		&	$	10036.48	\pm	1.23	$	&	$	26.30	\pm	4.94	$	\\
$	8675.02	\pm	0.30	$	&	$	86.79	\pm	4.47	$	&		&	$	10044.32	\pm	0.67	$	&	$	55.82	\pm	7.55	$	\\
$	8713.59	\pm	0.33	$	&	$	113.27	\pm	5.55	$	&		&	$	10154.46	\pm	0.44	$	&	$	99.85	\pm	5.49	$	\\
$	8765.54	\pm	0.42	$	&	$	42.84	\pm	3.09	$	&		&	$	10208.17	\pm	0.40	$	&	$	154.81	\pm	6.70	$	\\
$	8810.89	\pm	0.38	$	&	$	68.27	\pm	4.10	$	&		&	$	10356.43	\pm	0.44	$	&	$	98.20	\pm	5.05	$	\\
$	8833.81	\pm	0.62	$	&	$	19.05	\pm	1.96	$	&		&	$	10414.95	\pm	0.67	$	&	$	81.57	\pm	6.93	$	\\
$	8864.19	\pm	0.12	$	&	$	13.46	\pm	5.93	$	&		&	$	10443.78	\pm	0.40	$	&	$	212.08	\pm	10.08	$	\\
$	8905.57	\pm	1.25	$	&	$	10.06	\pm	1.84	$	&		&	$	10474.67	\pm	1.44	$	&	$	13.90	\pm	2.62	$	\\
$	8921.97	\pm	1.16	$	&	$	11.69	\pm	2.01	$	&		&	$	10511.30	\pm	1.00	$	&	$	10.38	\pm	3.22	$	\\
$	9000.02	\pm	0.31	^{*}$	&	$	89.86	\pm	4.79	$	&		&	$	10580.16	\pm	1.20	$	&	$	44.98	\pm	8.31	$	\\
 \hline
\end{tabular}

\end{table}

\newpage
\begin{table}[!htb]
\caption{Table \ref{kernels} continued.
\label{kernels5}}
\begin{tabular}{|c|c|c|c|c|c|}
\hline
$	E_R$ (eV)				&	$	k$ (meV)				&		&	$	E_R$ (eV)	&	$	k$ (meV)	\\ \hline
$	10598.22	\pm	1.10	$	&	$	138.91	\pm	18.94	$	&		&	$	11967.33	\pm	0.52	$	&	$	170.17	\pm	9.65	$	\\
$	10613.21	\pm	0.61	$	&	$	123.79	\pm	14.39	$	&		&	$	12030.65	\pm	0.90	$	&	$	31.06	\pm	3.08	$	\\
$	10683.28	\pm	0.61	$	&	$	86.73	\pm	6.60	$	&		&	$	12100.52	\pm	0.60	$	&	$	117.07	\pm	6.83	$	\\
$	10702.36	\pm	0.10	$	&	$	13.76	\pm	6.17	$	&		&	$	12180.40	\pm	1.15	$	&	$	35.23	\pm	3.62	$	\\
$	10746.07	\pm	0.71	$	&	$	112.63	\pm	11.42	$	&		&	$	12280.12	\pm	1.07	$	&	$	72.49	\pm	6.59	$	\\
$	10767.05	\pm	1.58	$	&	$	56.25	\pm	16.91	$	&		&	$	12323.72	\pm	0.66	$	&	$	69.06	\pm	5.94	$	\\
$	10776.24	\pm	1.12	$	&	$	233.72	\pm	29.00	$	&		&	$	12444.22	\pm	1.00	$	&	$	50.39	\pm	5.32	$	\\
$	10801.25	\pm	0.91	$	&	$	42.83	\pm	7.02	$	&		&	$	12558.97	\pm	0.63	$	&	$	113.80	\pm	7.86	$	\\
$	10851.47	\pm	0.42	$	&	$	155.87	\pm	8.83	$	&		&	$	12586.55	\pm	0.88	$	&	$	73.87	\pm	6.99	$	\\
$	10890.06	\pm	0.47	$	&	$	183.52	\pm	9.28	$	&		&	$	12664.81	\pm	0.42	$	&	$	146.51	\pm	8.08	$	\\
$	10945.27	\pm	0.69	$	&	$	33.74	\pm	3.13	$	&		&	$	12757.68	\pm	0.59	$	&	$	168.20	\pm	9.74	$	\\
$	11085.56	\pm	0.85	$	&	$	34.87	\pm	3.94	$	&		&	$	12849.82	\pm	0.55	$	&	$	236.74	\pm	10.65	$	\\
$	11136.11	\pm	0.81	$	&	$	68.18	\pm	8.26	$	&		&	$	12908.84	\pm	1.08	$	&	$	35.87	\pm	4.19	$	\\
$	11156.54	\pm	1.16	$	&	$	144.45	\pm	39.12	$	&		&	$	12979.04	\pm	0.96	$	&	$	52.03	\pm	6.60	$	\\
$	11161.78	\pm	1.08	$	&	$	178.91	\pm	40.17	$	&		&	$	13010.31	\pm	1.81	$	&	$	95.95	\pm	18.88	$	\\
$	11211.63	\pm	0.47	$	&	$	107.58	\pm	6.76	$	&		&	$	13029.01	\pm	1.35	$	&	$	107.14	\pm	18.96	$	\\
$	11240.54	\pm	1.05	$	&	$	19.73	\pm	2.91	$	&		&	$	13069.80	\pm	1.38	$	&	$	46.53	\pm	7.24	$	\\
$	11317.49	\pm	1.85	$	&	$	64.94	\pm	8.67	$	&		&	$	13113.37	\pm	1.32	$	&	$	28.32	\pm	3.89	$	\\
$	11343.57	\pm	0.88	$	&	$	58.18	\pm	7.70	$	&		&	$	13234.12	\pm	0.53	$	&	$	138.79	\pm	7.45	$	\\
$	11360.04	\pm	0.62	$	&	$	63.63	\pm	6.04	$	&		&	$	13289.11	\pm	0.65	$	&	$	74.27	\pm	11.67	$	\\
$	11409.51	\pm	1.03	$	&	$	40.34	\pm	4.50	$	&		&	$	13350.23	\pm	1.66	$	&	$	28.49	\pm	4.00	$	\\
$	11441.92	\pm	0.49	$	&	$	97.07	\pm	5.97	$	&		&	$	13384.20	\pm	1.26	$	&	$	50.88	\pm	6.13	$	\\
$	11482.53	\pm	1.55	$	&	$	16.56	\pm	3.21	$	&		&	$	13407.66	\pm	1.57	$	&	$	50.26	\pm	8.10	$	\\
$	11526.36	\pm	0.07	$	&	$	15.46	\pm	6.17	$	&		&	$	13425.24	\pm	0.78	$	&	$	121.41	\pm	11.21	$	\\
$	11543.91	\pm	0.17	$	&	$	14.97	\pm	6.42	$	&		&	$	13515.10	\pm	1.31	$	&	$	85.01	\pm	9.56	$	\\
$	11570.35	\pm	0.72	$	&	$	76.14	\pm	6.33	$	&		&	$	13551.33	\pm	0.52	$	&	$	280.04	\pm	14.04	$	\\
$	11597.57	\pm	0.98	$	&	$	40.01	\pm	4.31	$	&		&	$	13655.16	\pm	0.83	$	&	$	53.61	\pm	5.60	$	\\
$	11618.38	\pm	0.93	$	&	$	32.60	\pm	3.99	$	&		&	$	13700.20	\pm	0.68	$	&	$	102.99	\pm	7.16	$	\\
$	11675.00	\pm	0.68	$	&	$	116.44	\pm	9.37	$	&		&	$	13740.58	\pm	1.36	$	&	$	27.18	\pm	3.85	$	\\
$	11700.46	\pm	0.64	$	&	$	126.12	\pm	9.90	$	&		&	$	13767.27	\pm	1.49	$	&	$	32.44	\pm	4.85	$	\\
$	11795.03	\pm	0.64	$	&	$	118.44	\pm	7.41	$	&		&	$	13796.78	\pm	1.07	$	&	$	48.92	\pm	5.10	$	\\
$	11821.08	\pm	1.43	$	&	$	20.32	\pm	3.39	$	&		&	$	13857.56	\pm	0.76	$	&	$	133.77	\pm	10.78	$	\\
$	11855.97	\pm	1.58	$	&	$	21.54	\pm	3.56	$	&		&	$	13900.68	\pm	1.50	$	&	$	138.73	\pm	13.17	$	\\
$	11883.77	\pm	1.48	$	&	$	34.50	\pm	5.36	$	&		&	$	13940.97	\pm	1.11	$	&	$	88.29	\pm	10.83	$	\\
$	11911.03	\pm	0.45	$	&	$	188.90	\pm	10.26	$	&		&	$	13967.49	\pm	0.72	$	&	$	154.22	\pm	10.11	$	\\

 \hline

\end{tabular}

\end{table}

\end{document}